\documentclass{jpsj2}
\title{Internal Energy of the Potts model on the Triangular Lattice with 
Two- and Three-body Interactions}

\author{Masayuki Ohzeki and Hidetoshi Nishimori}

\inst{Department of Physics, Tokyo Institute of Technology,
Oh-okayama, Meguro-ku, Tokyo 152-8551}

\abst{We calculate the internal energy of the Potts model
on the triangular lattice with two- and three-body interactions
at the transition point satisfying certain conditions for coupling constants.
The method is a duality transformation.
Therefore we have to make assumptions on uniqueness of the transition point
and that the transition is of second order.
These assumptions have been verified to hold by numerical simulations for $q=2$, $3$ and $4$,
and our results for the internal energy are expected to be exact in these cases.
}

\kword{Potts model, exact energy, triangular lattice, duality transformation, transition point}
\begin{document}
\maketitle

\section{Introduction}
Exact solutions have exerted strong influences on the development of statistical physics
of equilibrium systems, the spin models in particular.
Among useful techniques for this purpose, one is the duality transformation,
which applies mainly to two-dimensional cases.
Ever since the first formulation of duality by Kramers and Wannier\cite{KW},
the method has been used to predict the transition points of the Ising model
on the square lattice\cite{KW}, the Potts model on the square,
\cite{Potts,Kihara} triangular, and hexagonal lattices\cite{KJ}.
The duality transformation makes it possible to obtain the location of
the transition point if it is unique.
We can also calculate the internal energy at the transition point
for a self-dual model from the derivative of the duality relation\cite{Kihara,Syozi}.

In the presented paper, we apply the duality transformation to the Potts model with
two- and three-body interactions on the triangular lattice and calculate
 the internal energy at the transition point.
The Potts model with two- and three-body interactions on the triangular lattice
was first introduced by Kim and Joseph\cite{KJ}.
They obtained this model as a result of an application of the duality and star-triangle
transformations to the Potts model only with two-body interactions on the triangular lattice.
Later, Baxter, Temperly and Ashley showed algebraically that the Potts model
with two- and three-body interactions is self-dual\cite{Baetal}, and so also did
Wu and Lin graphically\cite{Wu1}.
It was also shown that the Potts model with only three-body interactions on the triangular
lattice is self-dual\cite{Baetal}, though the Potts model
with only two-body interactions is not.

Baxter, Temperley and Ashley obtained the exact solution for the free energy and the
internal energy of the Potts model at the transition point with two-body interactions
on the triangular lattice\cite{Baetal}.
They used the relationship between this Potts model and the soluble Kelland model.
However, it was difficult to analyze the Potts model with both two- and three-body
interactions whereas the model is amenable to exact analysis by duality
if it has only three-body interactions\cite{Baetal}.
The Potts model with two- and three-body interactions has not been treated by
such considerations, because of the complexity of the relationship between the
original model and its dual although the duality for this model has been well known.

Our goal is to derive the exact solution of the internal energy at the transition point
for the Potts model with two- and three-body interactions using the duality transformation.
We find that the duality transformation gives us the value of the internal
energy only on some special points,
which lie on the line of transition points.
We also discuss the mechanism that we can calculate
the internal energy by the duality transformation.

The outline of the presented paper is as follows.
Section $2$ gives a brief introduction to the Potts model with two- and three-body 
interactions on the triangular lattice and its duality transformation.
Here we will find the condition that we can calculate the internal energy by use of
the duality transformation.
In \S $3$, we numerically verify the results obtained in \S $2$.
The last section is devoted to discussions.

\section{Energy of the Potts model with two- and three-body interactions}
\begin{figure}[tb]
\begin{center}
\scalebox{1.4}{\includegraphics*[85mm,130mm][130mm,170mm]{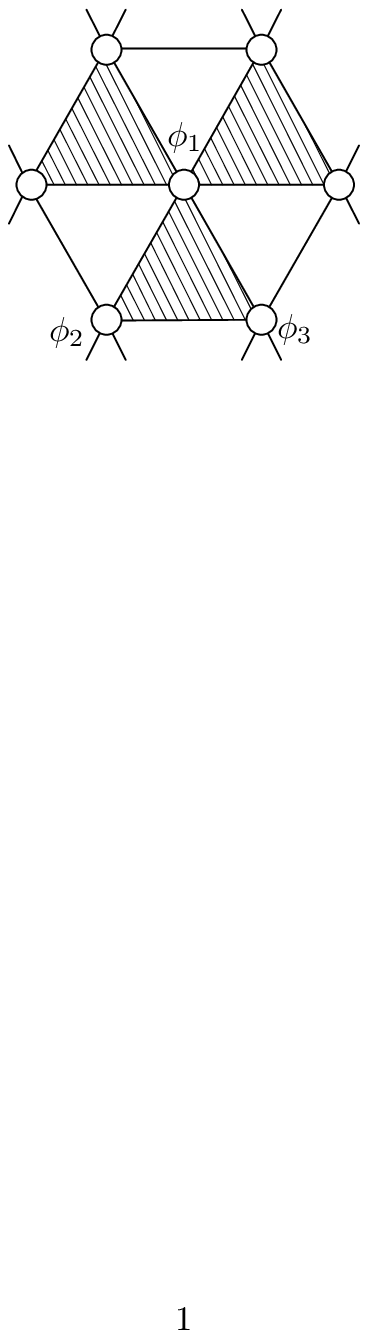}}
\end{center}
\caption{{\small The Potts model with two- and three-body interactions on the
triangular lattice. Each of the spin variables interacts with nearest-neighboring sites
(two-body interactions). We also consider three-body interactions among spins
on all up-pointing triangles as shown shaded.}
}
\label{tri}
\end{figure}

We introduce the $q$-state Potts model with two- and three-body interactions
on the triangular lattice and write its duality relations, following Ref. \citen{Wu2}).
The Hamiltonian is
\begin{equation}
H(J_2,J_3) \equiv \sum_{\bigtriangleup} H_{\bigtriangleup}(J_2,J_3) 
 = -J_2 \sum_{\bigtriangleup}\sum_{i \neq j} \delta \left( \phi_i,\phi_j \right)
 - J_3 \sum_{\bigtriangleup}\delta \left( \phi_1, \phi_2, \phi_3 \right),
\end{equation}
where \{$\phi_i$\} denotes the spin variables ($\phi_i=0,1,\cdots,q-1$).
The summation in the second expression is over all up-pointing triangles as shown in Fig. \ref{tri}.
Three-body interactions exist only on the up-pointing triangles.
Therefore our model is slightly different from the model of Schick and Griffiths\cite{SG},
which has three-body interactions on all up-pointing and down-pointing triangles.
The summation in the third expression with subscript $i \neq j$ is over bonds $(1,2)$,
$(2,3)$ and $(3,1)$ around each up-pointing triangle as shown in Fig. \ref{tri}.
The strengths of the two- and three-body interactions are denoted as $J_2$ and $J_3$,
and $\delta$ is the Kronecker's symbol.
The partition function is written as
\begin{equation}
Z(K_2,K_3) =
\sum_{\{\phi_i\}}\exp\left\{
-\beta H(J_2,J_3)
\right\} = \sum_{\{\phi_i\}} \prod_{\bigtriangleup} \exp 
\left\{
K_2 \sum_{i \neq j} \delta \left( \phi_i,\phi_j \right)
+ K_3 \delta \left( \phi_1, \phi_2, \phi_3 \right)
\right\},\label{def}
\end{equation}
where $K_2=\beta J_2$ and $K_3 = \beta J_3$.
This model is self-dual and the duality relation is known to be \cite{Baetal,Wu1,Wu2}
\begin{equation}
Z(K_2,K_3) = \left( \frac{y}{q} \right)^{N} Z(K^*_2,K^*_3),\label{dr}
\end{equation}
where $K^*_2$ and $K^*_3$ are the dual coupling constants and $N$ is the number of sites,
which is the same as the number of up-pointing triangles.
The relations between the coupling constants $(K_2,K_3)$ and $(K^*_2,K^*_3)$ are
\begin{equation}
v^* = \frac{qv}{y}, \quad y^* = \frac{q^2}{y},\label{PTRdr}
\end{equation}
where $v \equiv e^{K_2}-1$ and $y \equiv e^{K_3 + 3K_2}- 3 e^{K_2} + 2$
and similarly for $v^*$ and $y^*$ using $K^*_2$ and $K^*_3$.
The fixed-point condition, $K_2 = K^*_2$ and $K_3 = K^*_3$, is equivalent to $y=q$,
which describes a line in the $(K_{2},K_{3})$ plane.
It is expected that the system undergoes a phase transition across the line $y=q$. Indeed Wu and Zia argued that the line $y=q$ represents the critical surface if $J_3 + 3J_2 > 0$ and $J_3 + 2J_2 > 0$ \cite{Wu3}.

Next let us calculate the internal energy on the line $y =q$,
using the duality relation of the partition function (\ref{dr}).
In the case of a simple single-variable duality $Z(K)= c(K)^N Z(K^*)$, the logarithmic
derivative of both sides readily leads to the energy at the fixed-point,
\begin{equation}
 e(K_c) = -\frac{1}{c\left(1-\displaystyle\frac{dK^*}{dK}\right)}\left.
 \frac{dc}{dK}\right\rvert_{K=K_c},\label{duality_ene}
\end{equation}
where $e(K)$ is the energy per site, under the assumption of continuity $e(K_c+0)=e(K_c-0)$.
We therefore try to apply the same procedure to the present duality relation (\ref{dr})
with two variables.

The logarithmic derivative of the partition function with respect to $\beta$ gives
the internal energy per site as follows;
\begin{eqnarray}\nonumber
- \frac{1}{N}\frac{\partial}{\partial \beta} \ln Z(K_2,K_3)
&=& 
- \frac{J_2}{N}
\left\langle
\sum_{\bigtriangleup}
 \sum_{i \neq j} \delta \left( \phi_i,\phi_j \right)
\right\rangle_{K_2,K_3}
- \frac{J_3}{N}
\left\langle
\sum_{\bigtriangleup}
\delta \left( \phi_1, \phi_2, \phi_3 \right)
\right\rangle_{K_2,K_3}\\
&\equiv & J_2U_2(K_2,K_3) + J_3U_3(K_2,K_3),\label{ene}
\end{eqnarray}
where the brackets $\langle \cdots \rangle_{K_2,K_3}$ denote the thermal average.
The values of $U_2(K_2,K_3)$ and $U_3(K_2,K_3)$ on the line $y=q$ satisfy
the following equations, which are obtained by taking logarithmic derivatives of
eq. (\ref{dr}) by $K_2$ and $K_3$ and assuming continuity (i. e. second-order transition)
of these derivatives across the line $y=q$,
\begin{eqnarray}
  \left\{ 1- \left.\left(\frac{\partial K_2^*}{\partial K_2}
\right)\right\rvert_{y=q}\right\}U_2
      - \left.\left(\frac{\partial K_3^*}{\partial K_2}
  \right)\right\rvert_{y=q}U_3\label{ME1}
&=& \frac{1}{q}\left.\left(\frac{\partial y}{\partial K_2}
 \right)\right\rvert_{y=q}\\
- \left.\left(\frac{\partial K_2^*}{\partial K_3}\right)\right\rvert_{y=q}U_2
 +\left\{ 1- \left.\left(\frac{\partial K_3^*}{\partial K_3}
 \right)\right\rvert_{y=q}\right\}U_3
&=& \frac{1}{q}\left.\left(\frac{\partial y}{\partial K_3}
\right)\right\rvert_{y=q}.\label{ME2}
\end{eqnarray}
It turns out that these two equations are not independent of each other as
proved in Appendix\ref{determinant}.
Because of this property, we cannot calculate both $U_2$ and $U_3$ from eqs. (\ref{ME1}) and (\ref{ME2}), which makes it difficult in general to evaluate eq. (\ref{ene}).
Nevertheless, if the left-hand side of eq. (\ref{ME1}) is proportional to the final expression of eq. (\ref{ene}), eq. (\ref{ME1}) gives the internal energy apart from the proportionality constant (which can be calculated from the values of coefficients $U_2$ and $U_3$ in eq. (\ref{ME1})).
We therefore require that the ratio of coefficients of $U_2$
and $U_3$ in eq. (\ref{ME1}), $\{1-(\partial K_2^*/\partial K_2) \rvert_{y=q} \} / \{-(\partial K_3^*/\partial K_2) \rvert_{y=q}\},$ equal to $J_2/J_3$:
\begin{equation}
J_3\left\{1- \left.\left(\frac{\partial K_2^*}{\partial K_2}\right)\right\rvert_{y=q}\right\} 
 = -J_2\left.\left(\frac{\partial K_3^*}{\partial K_2}\right)\right\rvert_{y=q}\label{EEC}.
\end{equation}
We can write this condition explicitly in terms of $K_2$ and $K_3$, 
using the explicit forms of the derivatives in eqs. (\ref{deri1}) and (\ref{deri3})  in Appendix\ref{determinant}, as follows,
\begin{equation}
(J_3+3J_2)\frac{ v_c(6v_c + 3q)}{q(1+v_c)} = J_2\left[
3- \frac{1}{3v_c + q + 1}
\left\{3(1+v_c)-(3v_c+q)\frac{6v_c + 3q}{q}
\right\}
\right],
\end{equation}
where $v_c=e^{K_2}-1$ with $K_2$ being the value on the line $y=q$.
This equation is reduced to
\begin{equation}
(K_{3}+3K_{2})(3e^{K_{2}}+q-2)(e^{K_{2}}-1) = K_{2}(3e^{K_{2}}+2q-3)e^{K_{2}}.\label{EE}
\end{equation}
where $K_3$ is the value on the line $y=q$.
We can calculate the internal energy at the limited points on the line $y=q$ determined by eq. (\ref{EE}).

Let us emphasize once again that the condition (\ref{EE}) for the internal
energy to be calculable
has been obtained under the assumption of uniqueness and continuity of the transition.
The former uniqueness assumption was used in the identification of the fixed point $y=q$
with the transition point as discussed in Ref. \citen{Wu3}), and the latter continuity was necessary in deriving
eqs. (\ref{ME1}) and (\ref{ME2}).

\section{Results and numerical verification}

Equation ($\ref{EE}$) with $y=q$ has solutions for $q=2, 3$ and $4$
as well as for $q\ge 69$.
The results for the former case are listed in  Table \ref{Solutions}.
\begin{table}
\begin{tabular}{ccccc}
\hline
$q$ & $K_{2}$ & $K_{3}$ & $\tau$ & energy \\
\hline
$2$ & $0.375530$ & $0.347552$ & $0.925497$ & $-3.19412$ \\
$3$ & $0.291622$ & $0.737724$ & $2.52973$ & $-3.95336$ \\
$4$ & $-0.062177$ & $1.75913$ & $-28.2924$ & $-15.5884$ \\
$4$ & $-1.09861$ & $4.39445$ & $-4$ & $-0.5$ \\
\hline
\end{tabular}
\caption{{\small Solutions of eq. (\ref{EE}) and $y=q$. Here $\tau \equiv K_3/K_2$}.
}\label{Solutions}
\end{table}
The values of coupling constants in Table \ref{Solutions} satisfy the Wu-Zia criteria $J_3+3J_2>0$, and $J_3+2J_2>0$, which means that the existence of a transition in each case is very plausible.
We have checked the assumption of uniqueness and continuity of the transition
by numerical simulations for $q=2$, $3$ and $4$.
Systems with larger $q$ are expected to have first-order transitions since the
Potts model with two-body interactions on the square lattice has
first order transition for $q\ge 5$\cite{Wu2}.
We thus have not carried out numerical simulations for the case of larger $q$.

Most simulations have been performed for
linear system size $100$ and $10500$ MCS per spin.
The data have been averaged over $100$ independent runs.
The exception is the case with $q=4$ and $\tau =K_3/K_2= -4$,
in which we took $250$ independent runs of $4500$ MCS.

\begin{figure}[tb]
\begin{center}
\includegraphics[width=70mm,angle=270]{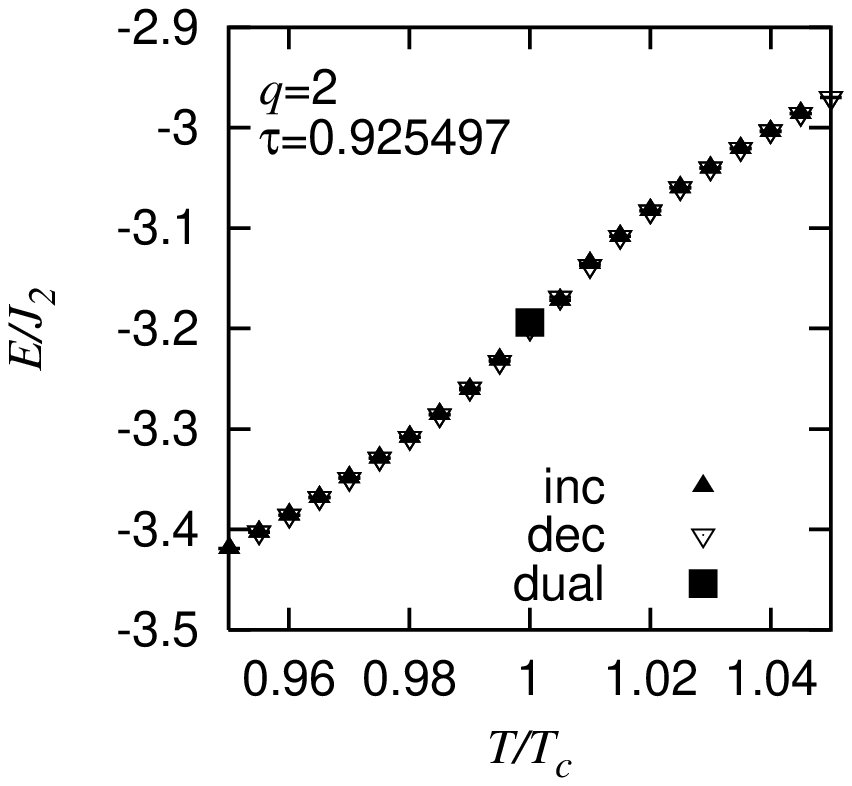}
\includegraphics[width=70mm,angle=270]{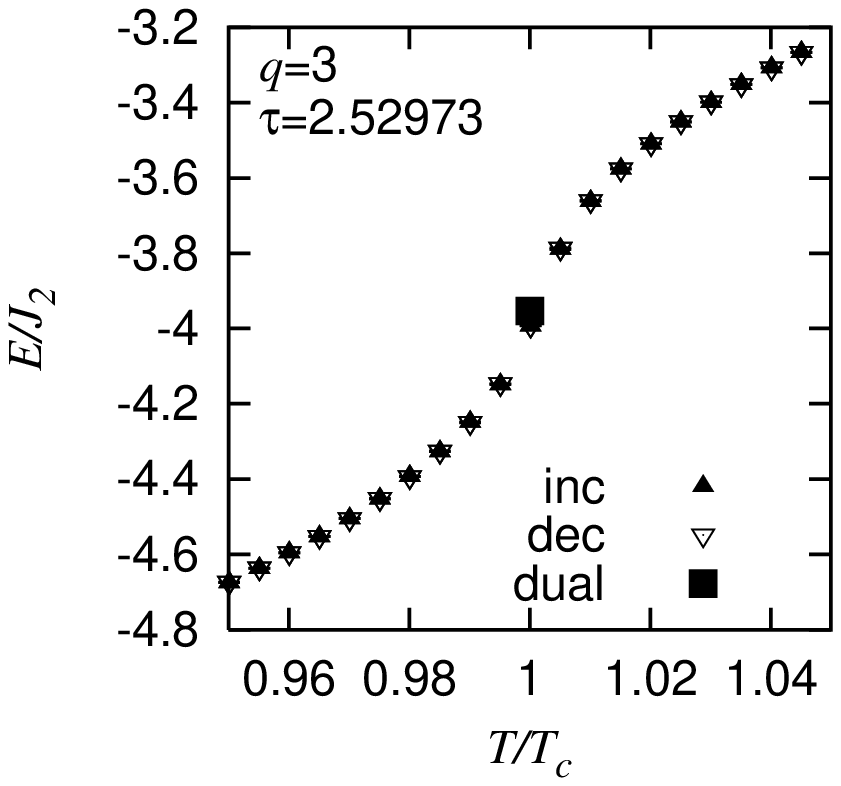}
\includegraphics[width=70mm,angle=270]{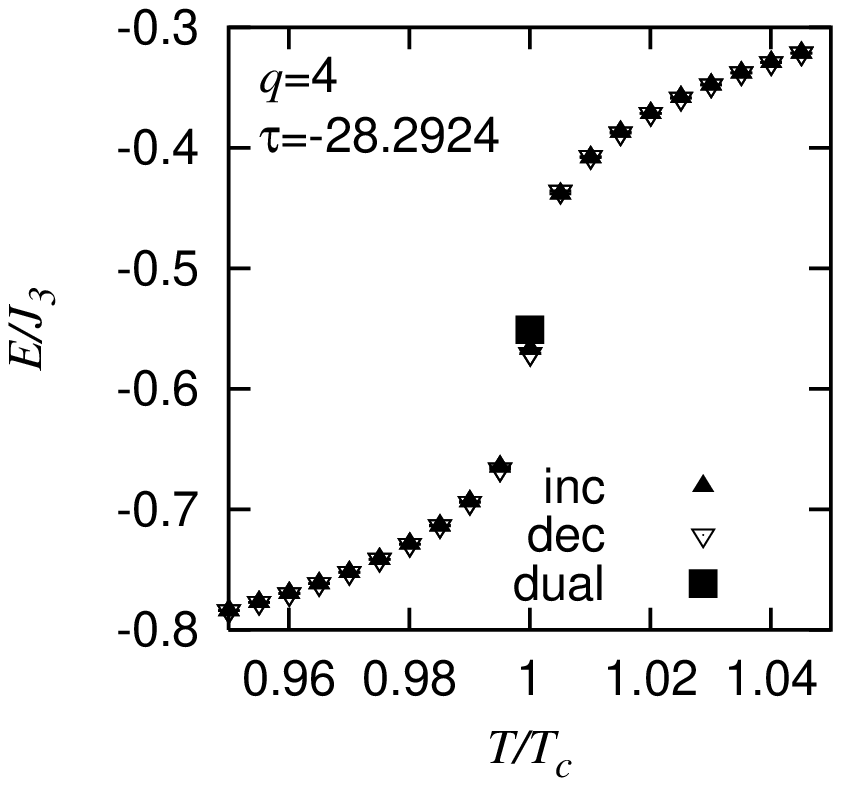}
\end{center}
\caption{{\small The internal energy at the fixed points of duality as in
Table. \ref{Solutions} except for $q=4$, $\tau=-4$.
The symbol `inc' denotes the internal energy measured by increasing the temperature,
`dec' stands for the results by decreasing temperature, and
`dual' is for our duality results.}
}
\label{NC}
\end{figure}

\begin{figure}[tb]
\begin{center}
\includegraphics[width=90mm,angle=270]{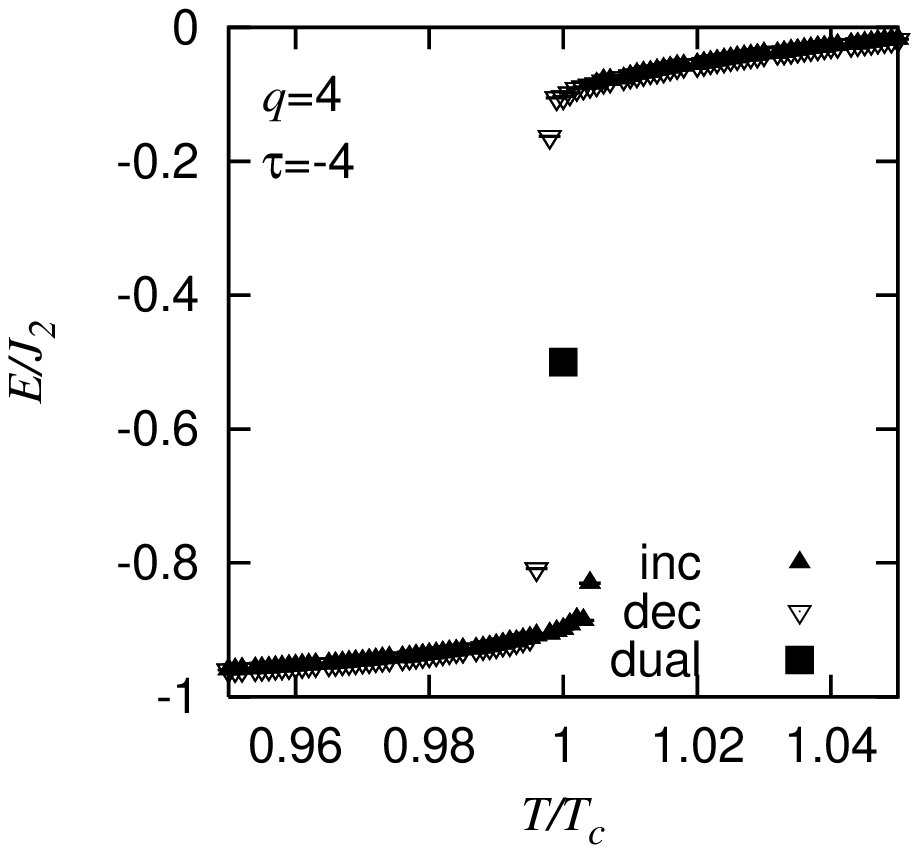}
\end{center}
\caption{
{\small The internal energy of the four-state Potts model with $\tau = -4$.
The same symbols are used as in Fig. \ref{NC}.}}
\label{44Potts1}
\end{figure}

The results are shown in Fig. \ref{NC} for $q=2$, $3$ and $4$ (with $\tau = -28.2924$)
and Fig. \ref{44Potts1} for $q=4$ with $\tau =-4$.
In Fig. \ref{NC} we observe that the internal energy shows no hysteresis,
which suggests that the assumption of continuity of transition is valid.
It is also seen that the duality predictions for the energy agree well with numerical data.
In contrast, the case with $q=4$ and $\tau=-4$ seems to undergo a first-order
transition as is seen in Fig. \ref{44Potts1}.
Thus the solution given in Table \ref{Solutions} for $q=4$ and $\tau=-4$ would correspond
to the average of $e(K_c+0)$ and $e(K_c-0)$, which is compatible with data in
Fig. \ref{44Potts1}: In the case of a simple self duality $Z(K)=c(K)^NZ(K^*)$,
taking the limit $K \to K_c +0$ after the logarithmic derivative leads to the average
of $e(K_c+0)$ and $e(K_c-0)$ on the left-hand side of eq. (\ref{duality_ene}).
The same situation is expected for $q=4$ and $\tau =-4$.
\section{Discussions}
We have calculated the internal energy of the Potts model on the
triangular lattice with two- and three-body interactions
satisfying certain conditions for coupling constants.
These results have been obtained from duality under the assumptions on uniqueness of
the transition point and that the transition is of second order.
We checked these assumptions by Monte Carlo simulations.
It has been shown that the assumptions are valid for $q=2$, $3$ and $4$ ($\tau=-28.2924$)
but not for $q=4$ ($\tau=-4$).
In the former case it is expected that the duality predictions are exact.
In the latter case our result gives the average energy
above and below the transition point.

\begin{figure}[tb]
\begin{center}
\includegraphics[width=70mm,angle=270]{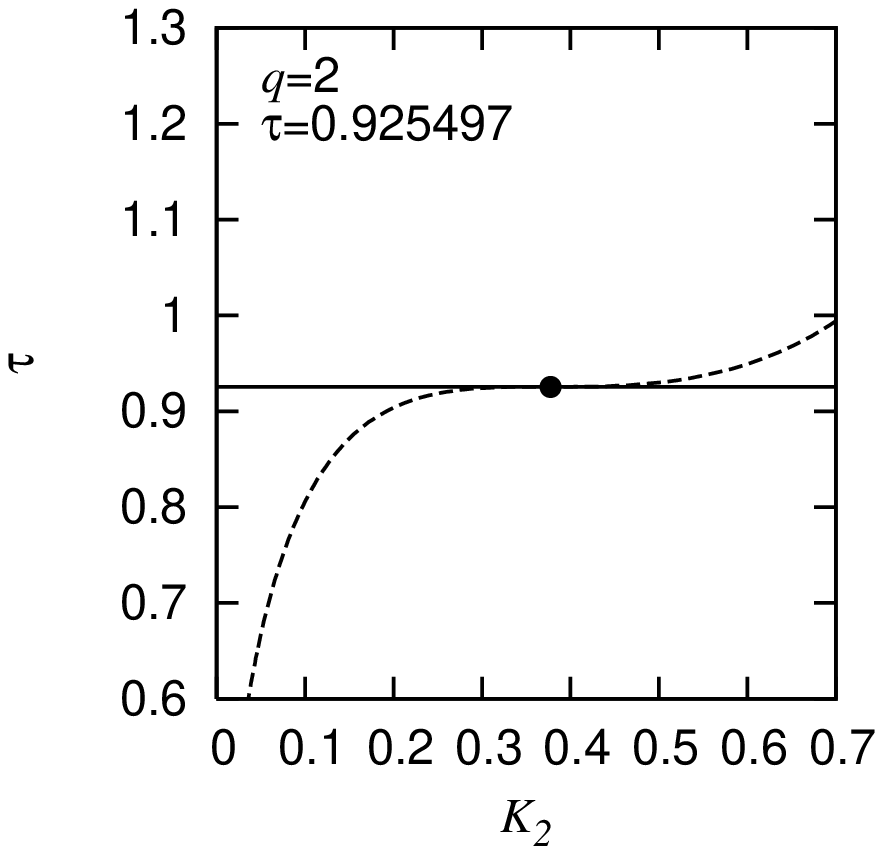}
\includegraphics[width=70mm,angle=270]{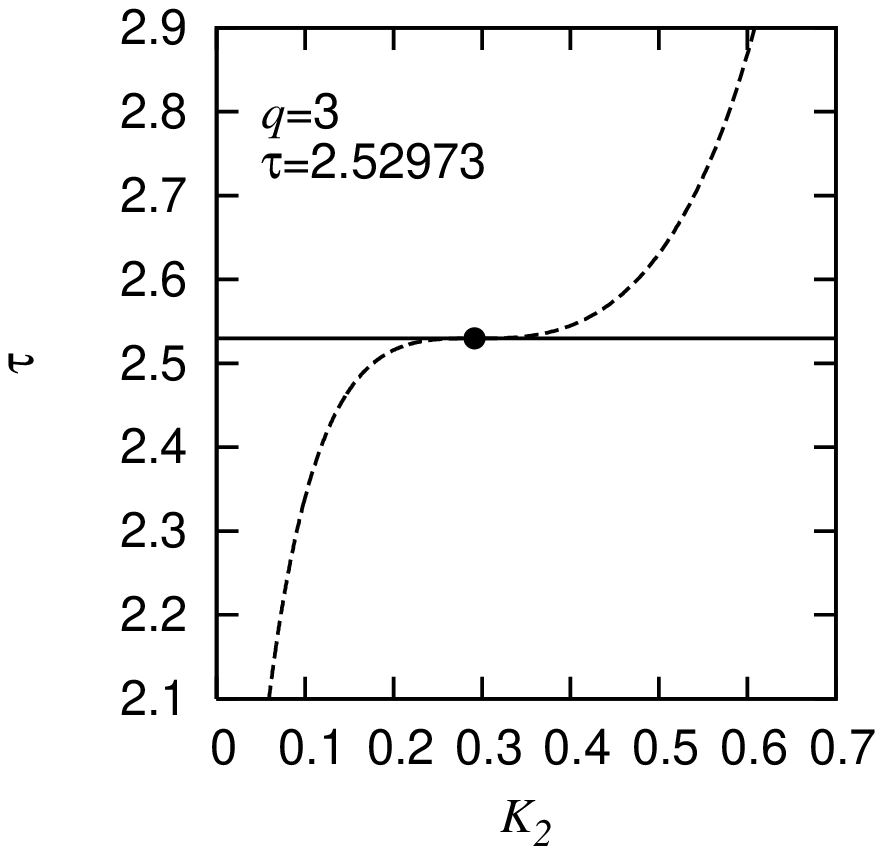}
\includegraphics[width=70mm,angle=270]{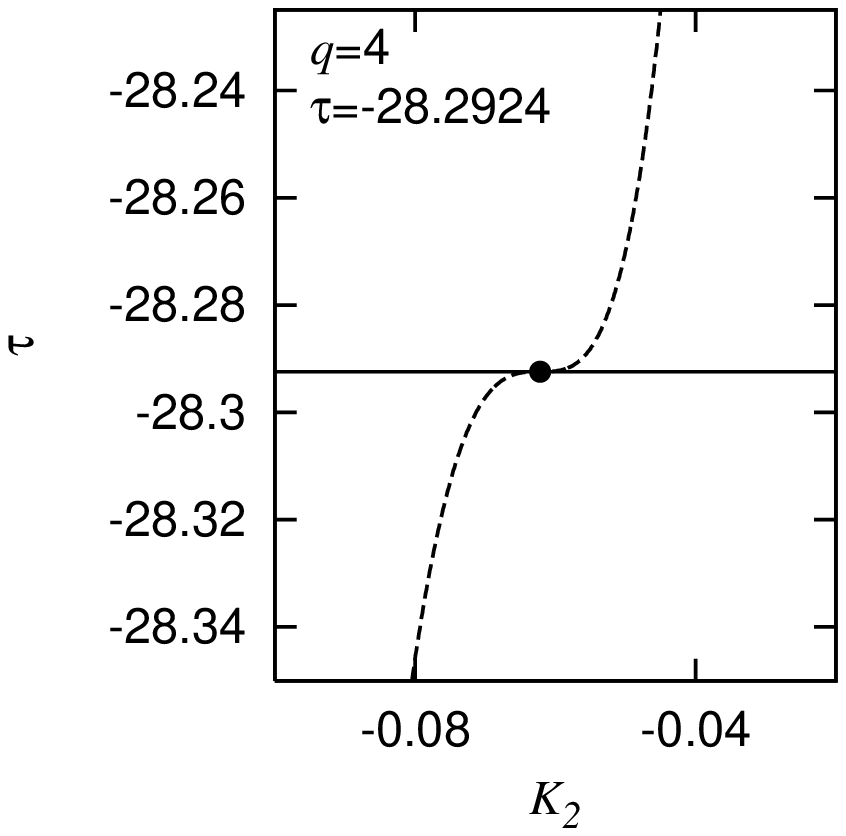}
\includegraphics[width=70mm,angle=270]{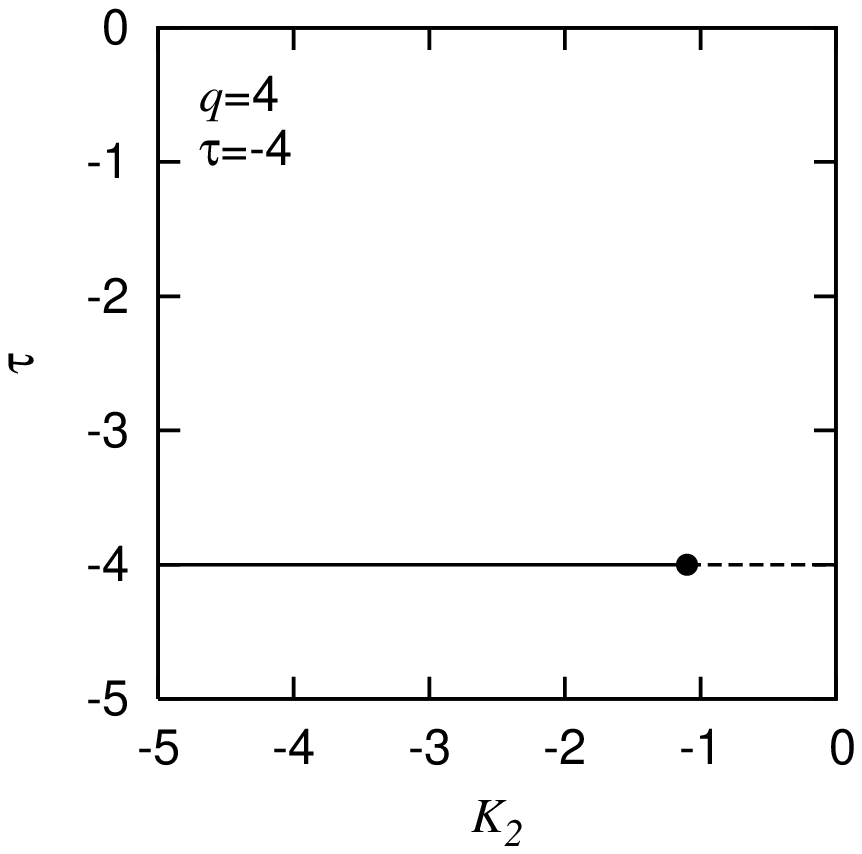}
\end{center}
\caption{{\small The behaviors around the solutions of eq. (\ref{EE}).
The horizontal solid line is the original model and the dashed line is for $K_3^*/K_2^*$.
These two lines intersect at the fixed point ($y=q$) shown in black dot.}
}
\label{Point}
\end{figure}

It is possible to understand the condition (\ref{EE}) for the internal energy to be
calculable from a little different point of view.
In Fig. \ref{Point}, $\tau=K_3/K_2$ and the dual $\tau^*=K^*_3/K^*_2$ are plotted as
functions of $K_2$.
As seen there, the two lines have the same slope (i.e. both are flat) at the fixed point.
This feature $dK^*_3/dK^*_2=dK_3/dK_2$ can be verified by direct manipulations of the
conditions (\ref{ME1})-(\ref{EE}).
In this sense the solvability condition may be regarded as a condition for the dual
model to be the same model as the original one, i.e., the same ratio of
coupling constants between two- and three-body interactions.
In other words the model is self-dual in a very strict sense only
at the transition point for $q=2$, $3$ and $4$ ($\tau=-28.2924$).

It is interesting to see the case of $q=4$, $\tau=-4$ in Fig. \ref{Point}: the model is
self-dual even away from the transition point. This is a very interesting
special feature of this case.
\section{Acknowledgement}
This work was partially supported by a 21st Century COE Program at Tokyo Institute of
Technology `Nanometer-Scale Quantum Physics' by the Ministry of Education, Culture,
Sports, Science and Technology and by CREST, JST.

\appendix
\section{Evaluation of the determinant}\label{determinant}
In this Appendix we show that eqs. (\ref{ME1}) and (\ref{ME2}) are not independent.
The dual coupling constants $K^*_2$ and $K^*_3$ are expressed by $y$ and $v$ as follows
\begin{eqnarray}
K^*_2 &=& \log \left( 1 + \frac{qv}{y} \right)\\
K^*_3 &=& -3K_2^* + \log\left( 1 + \frac{3qv+q^2}{y}\right),
\end{eqnarray}
where $y= e^{K_3+3K_2}-3e^{K_2}+2$ and $v = e^{K_2}-1$.
On the line $y=q$ , the derivatives of $y$ and $v$ are seen to satisfy
\begin{equation}
\frac{\partial y}{\partial K_2} = 6v_c + 3q,\quad
\frac{\partial y}{\partial K_3} = 3v_c + 1+ q,\quad
\frac{\partial v}{\partial K_2} = 1+v_c,\quad
\frac{\partial v}{\partial K_3} = 0,
\end{equation}
where $v_c \equiv e^{K_{2c}}-1$.
From these relations, we can obtain the derivatives of $K^*_2$ and $K^*_3$ with respect
to $K_2$ and $K_3$ on the $y=q$ line as follows
\begin{eqnarray}
\left.\frac{\partial K_2^*}{\partial K_2}\right\rvert_{y=q} &=&
1 - \frac{ v_c(6v_c + 3q)}{q(1+v_c)}\label{deri1}
\\
\left.\frac{\partial K_2^*}{\partial K_3}\right\rvert_{y=q} &=&
\frac{-v_c(3v_c + 1 + q)}{q (1 + v_c)}\label{deri2}
\\
\left.\frac{\partial K_3^*}{\partial K_2}\right\rvert_{y=q} &=&
-3\left.\frac{\partial K_2^*}{\partial K_2}\right\rvert_{y=q}
+ \frac{1}{3v_c + q + 1}
\left\{3(1+v_c)-(3v_c+q)\frac{6v_c + 3q}{q}
\right\}\label{deri3}
\\
\left.\frac{\partial K_3^*}{\partial K_3}\right\rvert_{y=q} &=&
-3\left.\frac{\partial K_2^*}{\partial K_3}\right\rvert_{y=q}
- \frac{(3v_c+q)}{q}.\label{deri4}
\end{eqnarray}
Using these derivatives, the determinant of eqs. (\ref{ME1}) and (\ref{ME2}) is evaluated as
\begin{eqnarray}\nonumber
&&\left(1- \left.\frac{\partial K_2^*}{\partial K_2}\right\rvert_{y=q}\right)
 \left(1- \left.\frac{\partial K_3^*}{\partial K_3}\right\rvert_{y=q}\right)
 - \left.\frac{\partial K_3^*}{\partial K_2}\right\rvert_{y=q} \left.
 \frac{\partial K_2^*}{\partial K_3}\right\rvert_{y=q}\\ \nonumber
&& = \frac{v_c(6v_c + 3q)}{q(1+v_c)}
 \left\{
 1+3\frac{-v_c(3v_c + 1 + q)}{q (1 + v_c)}- \frac{(3v_c+q)}{q}
 \right\}
\\ \nonumber
&& \quad
- \frac{-v_c(3v_c + 1 + q)}{q (1 + v_c)}
\left[-3
\left\{
1 - \frac{ v_c(6v_c + 3q)}{q(1+v_c)}
\right\}
+ \frac{1}{3v_c + q + 1}
\left\{3(1+v_c)-(3v_c+q)\frac{6v_c + 3q}{q}\right\}
\right]
\\ \nonumber
&& = 0.
\end{eqnarray}
Thus eqs. (\ref{ME1}) and (\ref{ME2}) are not independent.

\end{document}